\documentclass[prl,twocolumn,superscriptaddress, preprintnumbers]{revtex4}

\usepackage{amsmath}
\usepackage{amsfonts}

\usepackage{graphicx}
\usepackage{times}

\begin{document}

%%%%%%%%%%%%

\title{Effective Field Theory of Fractional Quantized Hall Nematics}

\preprint{MIT-CTP/4229}

\author{Michael Mulligan}
\affiliation{Center for Theoretical Physics, MIT, Cambridge, MA 02139, USA}

\author{Chetan Nayak}
\affiliation{Microsoft Station Q, Santa Barbara, CA 93106, USA }

\author{Shamit Kachru}
\affiliation{Department of Physics, Stanford University and SLAC,
Stanford, CA 94305, USA}

\begin{abstract}
We present a Landau-Ginzburg theory for a fractional
quantized Hall nematic state and the transition
to it from an isotropic fractional quantum Hall state.
This justifies Lifshitz-Chern-Simons
theory -- which is shown to be its dual -- on a more microscopic
basis and enables us to compute a ground state wave function
in the symmetry-broken phase. In such a state of matter,
the Hall resistance remains quantized while
the longitudinal DC resistivity due to thermally-excited
quasiparticles is anisotropic. We interpret recent experiments
at Landau level filling factor $\nu=7/3$
in terms of our theory.
\end{abstract}

\maketitle

%%%%%%%%%%%%

\paragraph{Introduction.}

A {\it fractional quantized Hall nematic} (FQHN) is a phase in which
a fractional quantized Hall conductance coexists
with the broken rotational symmetry characteristic
of a nematic, as in the model introduced
in Ref. \onlinecite{Mulligan10}.
The idea that a phase of matter could have both topological order
and conventional broken symmetry is not new; for instance,
quantum Hall ferromagnets are another example
\cite{Sondhi93,Barrett95}.  See \cite{ abanin} for a more recent discussion in a related system.
However, the FQHN has the unusual feature that 
the broken symmetry and the topological order
are equally important for determining the system's
transport properties. Furthermore, the model also predicts
an unusual quantum critical point separating the
FQHN from an ordinary isotropic fractional quantum Hall
state.

Remarkably, a recent experiment may have observed
a FQHN \cite{Xia11}.
An in-plane magnetic field $B_{\parallel}$ is applied to
the $\nu=7/3$ fractional quantum Hall plateau. When the
angle $\theta$ between the total magnetic field and the normal
is zero, the system is essentially
isotropic: for $T < 100$mK,
$R_{xx}\approx R_{yy}$. At $T = 15$ mK,
there is a well-developed Hall plateau with $R_{xy}=R_{yx}=\frac{3}{7}\,
\frac{h}{e^2}$. At $T>100$mK, there is a small ($\approx 20\%$)
difference between $R_{xx}$ and $R_{yy}$, which may be
due to device geometry, alignment of the contacts, or
a small intrinsic anisotropy acquired by the samples
during the growth process. For tilt angles $\theta>19^\circ$ and
$T<50$mK, $R_{xy}=R_{yx}=\frac{3}{7}\,\frac{h}{e^2}$ while 
$R_{xx}-R_{yy}$ increases with decreasing
temperature. In fact, $dR_{xx}/dT < 0$ while $dR_{yy}/dT > 0$
at the lowest observed temperatures. Thus, this experiment
finds transport which is reminiscent of the nematic phases found
at half-filling of higher Landau levels,
such as $\nu=9/2, 11/2, \ldots$ without an in-plane field \cite{lilly1, Fradkin10}
and also at $\nu=5/2$ and $7/2$ in the presence
of an in-plane field \cite{pan99, lilly2}, except for one very striking difference:
the Hall resistance remains quantized in the anisotropic phase.

We interpret these observations as a slightly rounded
transition between an isotropic fractional quantum Hall
phase at $\theta<\theta_c\stackrel{<}{\scriptscriptstyle \sim} 19^\circ$
and an FQHN at $\theta>\theta_c$.
The rounding of the transition is caused by the in-plane field.
We believe it to be a weak rotational symmetry-breaking field
because the system is in an isotropic metallic phase
for even larger tilts at the nearby fraction $\nu=5/2$ \cite{Xia10}
and because the anisotropy at 300 mK actually decreases as the
tilt is increased from $44^\circ$ to $76^\circ$. We conjecture that
the most important effect of the in-plane field is to vary the effective
interaction between the electrons, thereby driving the (almost)
spontaneous breaking of rotational symmetry. We are thus led
to apply our model \cite{Mulligan10} to this experiment.

To this end, we give a more microscopic derivation of our model
as a Landau-Ginzburg theory. We thereby recover a theory
which is equivalent, through
particle-vortex duality, to the effective field theory introduced in
Ref. \onlinecite{Mulligan10}. In order to compare theory and experiment
more closely, we extend our previous analysis of
zero-temperature, finite-frequency transport to finite-temperature DC transport;
in order to do this, we must enlarge our model to include the effects
of gapped charged quasiparticles.
The development of nematic order induces
strongly temperature-dependent anisotropy in the quasiparticle effective masses.
We predict that both longitudinal conductances will eventually
vanish at the lowest temperatures, although one of them will
have non-monotonic temperature dependence at slightly higher temperatures.
We finally make predictions for transport at and near
the transition point.

\paragraph{Landau-Ginzburg Theory.}

One can map the problem of spinless planar electrons
in a transverse magnetic field $B$ with Coulomb repulsion,
to an equivalent system of a bosonic order parameter $\phi$ of unit charge coupled to a Chern-Simons gauge field $a_\mu$ \cite{ZHK}.  The
action takes the form:
\begin{eqnarray}
\label{CSLG}
 S_{LG}  & = &  \int d^2 x dt\Bigl(\phi^\dagger i(\partial_t - i(A_t + a_t))\phi \cr
& - & {1\over 2m_e}
\vert (\partial_i - i(A_i + a_i))\phi\vert^2 
+ {\nu \over 4 \pi} \epsilon_{\alpha\beta\gamma} a_{\alpha} \partial_{\beta}
a_{\gamma}  \cr
 & - & {1\over 2} \int d^2 y (\phi^\dagger \phi(x) - \bar\rho) V(x-y) (\phi^\dagger\phi - \bar\rho)\Bigr).\
\end{eqnarray}
$A_\mu$ is the background electromagnetic field satisfying
$\epsilon_{ij} \partial_i A_j = B$; 
$\bar\rho$ is the mean charge density of bosons (or equivalently, electrons);
$m_e$ is the electron band mass; V(x) is a general two-body potential; and
the Chern-Simons gauge field $a_\mu$ attaches $2 \pi \nu^{-1}$ units of 
statistical flux to each particle \cite{CS-normalization}.
In particular, for $\nu^{-1}$ an odd integer, the resulting
Aharonov-Bohm phases transmute the bosons
into fermions.

We assume that the low-energy effective theory
for distances longer than the magnetic length,
obtained by integrating out short-distance fluctuations of
$\phi$, $a_\mu$, has the same form as the microscopic action (\ref{CSLG}),
but with the bare microscopic parameters $1/m_e$ and $V(x-y)$
replaced by renormalized ones, $\bar{r}$ and $V_{\rm eff}(x-y)$.
Such an ansatz allows one to derive many of the properties of the standard
fractional quantum Hall states \cite{ZHK, ZhangIJMP}.  
Here, we will make the same ansatz, but without
assuming that $\bar{r}$ remains positive.
We note that even the `microscopic' action (\ref{CSLG})
must be viewed as an effective low-energy action that describes the partially filled $N=1$ Landau level with $\nu = 2+1/3$. The electrons are
confined to a quantum well of finite-width; a strictly two-dimensional
theory is an effective theory at energy scales far below the splitting
between energy sub-bands for motion perpendicular to the plane.
Thus, the application of the in-plane field $B_{||}$, through its modification
of the motion perpendicular to the plane, will modify the parameters
in $S_{LG}$.
Consequently, the effective parameters at distances longer than the
magnetic length will also be modified, but not in a simple or, at present,
transparent way.
It is easy to check that reasonable local variations of $V_{\rm eff}$
do not cause qualitative changes to the physics of (\ref{CSLG}) \cite{ZhangIJMP}. 
We leave to a future study the question of higher-body potential terms
resulting from a projection of the degrees of freedom into a specific Landau level.

Therefore, we conjecture that as the in-plane field $B_{||}$ is varied, the most significant variation is of the parameter
$\bar{r}$. 
In other words, we study the instabilities of (\ref{CSLG}) as the kinetic structure of the  theory is modified. 

Since we will be considering $\bar{r}<0$,
we add the following term with $c>0$ to the action in order
to maintain stability of the vacuum:
\begin{equation}
\label{CSLGmod}
\delta S = - {c \over 2} \int d^2x dt  \vert(\partial_i - i (A_i + a_i))^2 \phi\vert^2~.
\end{equation}  

This theory exhibits a transition between an isotropic fractional quantum Hall phase, when $\bar{r}>0$, and an anisotropic phase
with well-quantized Hall conductance (after inclusion of disorder or
a lattice) when $\bar{r}<0$, just as in \cite{Mulligan10}.  The two phases
are separated by a quantum critical point with $z=2$ dynamical scaling,
arising at $\bar{r}=0$.

\paragraph{Kohn's Theorem.}

On might object to any variation of $\bar{r}$ from its bare value on
the basis of Kohn's theorem \cite{kohn}. (See Section 5 of \cite{ZhangIJMP} for a discussion.)
In a Galilean-invariant system of $N$ identical mutually interacting particles
of unit charge and mass $m_e$ subject to a constant external magnetic field $B$,
Kohn's theorem states that the density-density correlation function has the low momentum limit,
\begin{equation}
\label{density}
\lim_{q \rightarrow 0} {\langle \rho(\omega, q) \rho(-\omega, -q) \rangle \over q^2} = {{1 \over m_e} \over \omega^2 - \omega_c^2}, 
\end{equation}
The locations of the two poles are determined by the cyclotron frequency $\omega_c = B/m_e$.
For fixed $B$, the cyclotron frequency is determined by the bare mass of the particles, independent of the their relative interactions.
The residues of the poles are equal to $\pm 1/2 B$.  
The form of this correlator is ensured by a Ward identity and satisfies an f-sum rule.  
Kohn's theorem roughly states that the center-of-mass of the system always decouples from the relative coordinate motion of the particles; it effectively behaves as a single charge
$N$ particle of mass $N m_e$, exhibiting circular motion at a frequency $\omega_c$
in a background magnetic field $B$. In quantum Hall systems, the quantum well
explicitly breaks translational symmetry in the $z$-direction (i.e. perpendicular to the plane).
However, the in-plane center-of-mass motion still decouples from the other degrees
of freedom, so long as the magnetic field is strictly perpendicular to the plane.
Thus, Kohn's theorem holds even in this case.

If we now compute the density-density correlator using the action (\ref{CSLG}),
we find precisely the form dictated by Kohn's theorem (\ref{density}).
However, the modification $1/m_e \rightarrow \bar{r}$ would change the location of the pole.
This manifestly constitutes a violation of Kohn's theorem.  

However, the experiment of Ref. \onlinecite{Xia11} does not
satisfy the assumptions of Kohn's theorem.
The large in-plane field, combined with the confining well potential (perpendicular to the plane),
manifestly breaks Galilean invariance and does not allow a decoupling of the center-of-mass mode.
The in-plane field couples motion along the $z$-direction to motion in the plane,
while the confining potential in the $z$-direction couples the $z$-component of the
center-of-mass position to the $z$-component of the relative coordinates.
The $N=1$ Landau level in the devices considered in Ref. \onlinecite{Xia11}
is particularly susceptible to perturbations mixing planar and $z$-direction
motion because the gap to the $N=0$ Landau level of the next quantum well
sub-band is small \cite{Xia10}.

In summary, our theory, in which $\bar{r}$ is not fixed, applies to situations,
such as those in the experiment of Ref. \onlinecite{Xia11}, in which Kohn's theorem
does not hold. Our theory cannot describe a fictional system in which the two-dimensional
layer is infinitely-thin and the transition is driven purely by tuning the inter-electron
interaction (without any in-plane field) since such a system would
necessarily satisfy Kohn's theorem.
To make our point more concrete, we show in the Appendix
that, as a result of the violation of the conditions of Kohn's theorem, the location of the
cyclotron pole can vary as $B_{||}$ is increased.

\paragraph{Duality.}

We have computed the long wavelength transport properties of the various phases  of (\ref{CSLG}) directly from the Landau-Ginzburg theory and found them to exactly match the response determined from the Lifshitz-Chern-Simons (LCS) theory of \cite{Mulligan10}.  This is expected because there is a low-energy equivalence between the (more) microscopic theory (\ref{CSLG}), (\ref{CSLGmod}) and the LCS theory which we demonstrate by expanding about the relevant ground state
in the three cases $\bar{r}>0, \bar{r}=0, \bar{r}<0$, and mapping the low-energy theory to the action governing the similar phase of the LCS theory, using particle-vortex duality \cite{fisherlee1}. For convenience, we assume a short-ranged repulsive interaction, $V_{\rm eff}(x) = V_0 \delta(x)$ with
$V_0 > 0$, throughout. 
This choice is motivated by expected screening effects of the microscopic electrons.
Nevertheless, the precise form of $V_{\rm eff}$ plays very little role in the considerations below as long as it is local.

For $\bar{r} \geq 0$,
there is a saddle point configuration given by
$\langle \phi^{\dagger} \phi \rangle = \bar{\rho}$,
$\langle a_\mu \rangle = - A_\mu$,
with filling fraction $\bar{\rho}/B = \nu /2\pi$.   
The low-energy action for fluctuations about this ground state when $\bar{r} > 0$ is
\begin{eqnarray}
 S_{\rm eff}(\bar{r}>0) & = &  \int d^2x dt \Bigl(-\delta\rho (\partial_t \theta - \delta a_t)  - {\bar{r} \over 2} \bar\rho (\partial_i \theta - 
\delta a_i)^2 \cr
& + & {\nu \over 4 \pi} \epsilon_{\alpha\beta\gamma}\delta a_{\alpha} \partial_{\beta}\delta a_{\gamma} 
- {1\over 2} V_0 (\delta \rho)^2\Bigr).
\end{eqnarray}
$\delta \rho$ and $\theta$ govern the fluctuations of the norm and phase of the bosonic order parameter $\phi$, $\delta a_\mu$ represents the fluctuation of the Chern-Simons gauge field, and we have taken the background field fluctuations to vanish.
$S_{\rm eff}(\bar{r}>0)$ can be rewritten by introducing the field $J_i$ (the spatial components of the $U(1)$ current associated with
the background gauge field):
\begin{eqnarray}
S_{\rm eff}(\bar{r}>0) & = & \int d^2x dt \Bigl(- \delta\rho(\partial_t \theta - \delta a_t) - J_i (\partial_i \theta - \delta a_i) \cr
& + &   {1\over 2\bar{r} \bar\rho} J_i^2 + { \nu \over 4 \pi} \epsilon_{\alpha\beta\gamma}\delta a_{\alpha}\partial_{\beta}
\delta a_{\gamma} - {1\over 2} V_0 \delta\rho^2\Bigr).
\end{eqnarray}
Now, integrating out $J_i$ trivially reproduces the previous Lagrangian; but we can instead find a dual description of
the theory by keeping $J_i$ in the Lagrangian and integrating out the other degrees of freedom.  $\theta$ appears
linearly and functions as a Lagrange multiplier ensuring conservation of $J_\mu$.  We can guarantee this by rewriting
$J_\mu = {1 \over 2  \pi} \epsilon_{\mu\nu\tau} \partial_{\nu} n_{\tau}$.  Writing the theory in terms of $n$, and 
integrating out $\delta a_{\mu}$, we find
\begin{eqnarray}
S_{LCS}(\bar{r}>0)  =   \int d^2x dt \Bigl({1\over 2g_e^2}(\partial_i n_t - \partial_t n_i)^2 \cr
 -  {1\over 2g_m^2} (\partial_i n_j - \partial_j n_i)^2 + {1\over 4\pi \nu} \epsilon_{\alpha\beta\gamma} n_{\alpha}\partial_{\beta}n_{\gamma}\Bigr).
\end{eqnarray}
This is Maxwell-Chern-Simons theory at level $\nu^{-1}$ with $g_e^2 = 4\pi^2 \bar{r} \bar\rho$ and $g_m^2 = {4\pi^2 \over  V_0}$.   This
matches the behavior of the LCS theory of \cite{Mulligan10} in the fractional quantum Hall phase $(\bar{r}>0)$.

When $\bar{r} = 0$ (the $z=2$ critical point), it is necessary to keep the $\delta S$ term.  Nevertheless, the dualization proceeds almost identically.  The leading terms in 
the expansion of the action in small fluctuations about the saddle point are
\begin{multline}
S_{\rm eff}(\bar{r}=0)  =   \int d^2x dt \Bigl( - \delta\rho(\partial_t \theta - \delta a_t) - {J_i \over \partial^2} \bigl(\partial_i \partial_j (\partial_j \theta \cr
 -  \delta a_j) -  {1\over 2 c\bar\rho} J_i \bigr) + { \nu \over 4 \pi} \epsilon_{\alpha\beta\gamma} \delta a_{\alpha} \partial_{\beta} \delta a_{\gamma} - {1\over 2} V_0 \delta\rho^2\Bigr).
\end{multline}
This is a formal expression because of the inverse Laplacian in the second term.  
Current conservation, which is imposed by the $\theta$ equation of motion,
allows us to replace $J$ with the emergent gauge field $n$.    
Imposing the gauge conditions $\delta n_0 = 0$ and $\partial_i n_i = 0$, and integrating out $a_{\mu}$,
we obtain a gauge-fixed version of the LCS Lagrangian.  Covariantizing the gauge-fixed action yields
\begin{eqnarray}
S_{LCS}(\bar{r}=0) =  {1 \over g^2} \int d^2x dt \Bigl({1\over 2 \kappa^2} {1\over \partial^2} (\partial_i n_t - \partial_t n_i)^2 \cr
 - {1 \over 2} (\partial_i n_j - \partial_j n_i)^2 + {g^2 \over 4\pi \nu} \epsilon_{\alpha\beta\gamma} n_{\alpha}\partial_{\beta}
n_{\gamma}\Bigr),
\end{eqnarray}
where $\kappa^2 = 2 c \bar{\rho} V_0$, and $g^2 = 4\pi^2/V_0$.  This is precisely the theory governing the critical point in \cite{Mulligan10}, with the $e_i$ field integrated out. (The $z=2$ nature of the $e_i$ field action $\sim (\partial_i e_j)^2$
in that theory, gives rise to the peculiar inverse Laplacian in the action above).

Lastly, we discuss the anisotropic $\bar{r}<0$ phase.  
The ground state is still homogeneous,
$\langle \phi^\dagger \phi \rangle = \rho^\prime$
$\rho^\prime = \bar\rho + { |\bar{r}|^2}/{8 cV_0}$,
but anisotropic, since $\langle a_\mu \rangle = - A_\mu - v_\mu$,
with, $v_0=0$ and $v_i^2 = {|\bar{r}|}/{2c}$.
At this saddle point, the chemical potential is shifted upwards.
The leading terms in the low-energy
action, expanding around the symmetry-breaking vacuum with the condensate lying along the x-axis, take the form (where again we have introduced a current $J_i$)
\begin{eqnarray}
\label{eqn:r<0-eff-action}
S_{\rm eff}(\bar{r}<0)  & = & \int d^2 x dt \Bigr(-\delta\rho (\partial_t \theta - \delta a_t)
-  J_x [(\partial_x \theta - \delta a_x) \cr
& - & {1\over 4 |\bar{r}|\rho^\prime}J_x^2)] 
-  {J_y\over \partial_y^2} [\partial_y^2(\partial_y\theta - \delta a_y) - {1\over 2 c\rho^\prime}J_y^2] \cr
& + &  {\nu \over 4 \pi} \epsilon_{\alpha\beta\gamma}\delta a_{\alpha}\partial_{\beta}\delta a_{\gamma} - {1\over 2} V_0 \delta \rho^2\Bigr).
\end{eqnarray}
The $\theta$ equation of motion imposes current conservation for the density $\delta\rho$ and current $J_i$.  
Integrating out $\delta a_\mu$  once more, we obtain
\begin{multline}
S_{LCS}(\bar{r}<0)  =  {1 \over g^2} \int d^2x dt \Bigl( {1\over 2 \kappa^2}{1\over \partial^2}(\partial_x n_t - \partial_t n_x)^2 \\
+ {g^2 \over 2|r|}(\partial_y n_t - \partial_t n_y)^2 -
{1\over 2}(\partial_i n_j - \partial_j n_i)^2
+ {g^2 \over 4\pi \nu}\epsilon_{\alpha\beta\gamma}n_{\alpha}
\partial_{\beta}n_{\gamma} \Bigr),
\end{multline}
where $\kappa^2 = 2 c \rho' V_0$, $|r| = 4 |\bar{r}| \rho' V_0$, and $g^2$ is as above.  This agrees with the LCS theory in the anisotropic phase in \cite{Mulligan10}. It is gapless, as may be seen from the $n_i$
propagators, which evince a contribution from the Goldstone mode
for spontaneously-broken SO(2) rotational symmetry. 
Note that a symmetry breaking vacuum
along the x-direction of the LG theory corresponds to a symmetry breaking vacuum along the y-direction in the LCS theory.

The effects of disorder are implemented by allowing spatially varying
$\bar{r}(x)$ in the Landau-Ginzburg description.
The low-energy equivalence implies that introducing such disorder
in the LG theory will 
lift the Goldstone mode of the spontaneously broken SO(2) symmetry
and will lead to a quantized Hall conductance,
as it did in in the anisotropic phase of the LCS theory \cite{Mulligan10}.
The pseudo-Goldstone mode should be visible in low-energy Raman scattering experiments.
Alternatively, we could introduce a lattice by including terms in
the action which explicitly lower the rotational symmetry from
SO(2) to $D_4$. In this case, the third term in 
(\ref{eqn:r<0-eff-action}) takes, instead, the form
$J_y [(\partial_y \theta - \delta a_y) - {1\over 4 |\bar{r'}|\rho^\prime}J_y^2)]$,
where $\bar{r}'$ is proportional to the effective lattice potential;
consequently, there is no Goldstone mode for
rotational symmetry-breaking.

\paragraph{Ground State Wave Function in the $\bar{r}<0$ Phase.}

We now compute the ground state wave function in the
 $\bar{r}<0$ phase following the method described in \cite{ZhangIJMP}.
For $D_4$ symmetry, which is more experimentally-relevant,
it takes the form:
\begin{equation}
\label{wavefunction}
\Psi(z_i) = \prod_{i<j} (z_i - z_j)^{1/v} \left(1 + \mbox{${\delta \bar{r} \over |\bar{r}| \nu}
{(z_i - z_j)^2 + (\bar{z}_i - \bar{z}_j)^2 \over |z_i - z_j|^2}$}\right).
\end{equation}
In (\ref{wavefunction}), $z_i=x_i + i y_i$, $\delta \bar{r}=\bar{r}-\bar{r}'$, and
we have suppressed both higher-order terms in $\delta r/r$ and the
$\exp(- \sum_i |z_i|^2/4 \ell_0^2)$ Gaussian factor where $\ell_0^2 = \hbar/B$ .  
The wave function becomes identical to the Laughlin wave function
in the absence of symmetry breaking, $\delta \bar{r} = 0$.
It would be interesting to understand if there is any relation
between ($\ref{wavefunction}$) and Ref. \onlinecite{joynt}.

\paragraph{Finite-temperature Transport.}

We now compute the contribution to the
finite temperature DC conductivity tensor from thermally-excited
charged quasiparticles. The LCS theory is more convenient than the equivalent
Landau-Ginzburg description because (massive) charged quasiparticles
are vortices of the Landau-Ginzburg theory and fundamental particles of the LCS theory. This computation demonstrates
that highly-anisotropic finite-temperature transport can result
from our model but is not an attempt to give a precise fit
to experimental data, which would require a more careful analysis
of the effects of disorder, the lattice, and subleading interactions.

We include the effects of the massive quasiparticles by adding to the `first-order' form of the LCS action,
\begin{eqnarray}
S_{LCS} & = &  {1 \over g^2} \int d^2x dt \Bigl(e_i \partial_t n_i + n_t \partial_i e_i - {r \over 2} e_i^2 
-  {\kappa^2 \over 2} (\partial_i e_j)^2 \cr 
& - & {1 \over 2} (\epsilon_{ij} \partial_i n_j)^2 +  {g^2 \over 4 \pi \nu} \epsilon_{\mu \nu \lambda} n_\mu \partial_\nu n_\lambda 
 - {\lambda \over 4} (e_i^2)^2 \cr 
& + & {\alpha \over 4} (e_x^4 + e_y^4) + {1 \over 2\pi} \epsilon_{\mu \nu \lambda} A_\mu \partial_\nu n_\lambda \Bigr),
\end{eqnarray}
the matter action,
\begin{eqnarray}
S_{\rm matter} & = &  \int d^2x dt \Phi^* \Bigl(i\partial_t + n_t - \Delta + (i\partial_i + n_i)^2 \cr
 & + & u\ {e_x^2} \, (i\partial_x + n_x)^2 + u\ {e_y^2}\,  (i\partial_y + n_y)^2\Bigr) \Phi.
\end{eqnarray}
Thus, we study the total action $S = S_{LCS} + S_{\rm matter}$.  In $S_{LCS}$, we have not integrated out the $e_i$ field.  
At tree-level, the quartic $e^4$ terms in $S_{LCS}$ are marginal; the operator with coefficient $\lambda$ preserves the full spatial $SO(2)$ symmetry, while the operator with coefficient $\alpha$ explicitly breaks it down to $D_4$.
We assume $\alpha$ is small and positive, reflecting a small explicit breaking of
$SO(2)$ inherent in the real material.
The last term in $S_{LCS}$ is the coupling to the external electromagnetic field $A_\mu$.
The statistical gauge field endows the massive quasiparticles represented by $\Phi$
with their fractional statistics. The irrelevant energy-energy coupling parameterized by $u$ is the leading term that directly communicates the $D_4$ spatial rotational symmetry breaking of the $r<0$ ground state to the matter field. By ignoring a possible $e_i^2 |\Phi|^2$ coupling, we are assuming that the magnitude of the symmetry-breaking order parameter $\langle e_i \rangle$ in the $r<0$ regime is much less than the quasiparticle gap $\Delta$.

We concentrate on the finite temperature DC conductivity when $r<0$, however, the actual expressions obtained are valid for all $r$, if interpreted appropriately.  
(The functional form of the optical conductivity was already determined in \cite{Mulligan10}; it differs in the two phases, and shows striking features at the critical point.)  
Let us assume that $\langle e_x \rangle$ is non-zero in the $r<0$ regime at zero temperature.  To quadratic order, $S_{\rm matter}$ becomes
\begin{multline}
S_{\rm matter} = \int d^2 x dt \Bigl(\Phi^*(i \partial_t + n_0 - \Delta)\Phi \\
+ \Phi^*((1 + u \langle e_x \rangle^2)(i \partial_x + n_x)^2 + (i \partial_y + n_y)^2 )\Phi\Bigr).
\end{multline}

At temperatures less than $\Delta$, we can integrate out the quasiparticles and write an effective action solely in terms of the fields appearing in $S_{LCS}$.  It is convenient to express the resulting effective action in Fourier space, obtaining
\begin{multline}
\label{eqn:eff-Action-LCS+A}
S = S_{LCS} + {1 \over 2} \int \! d^2q d\omega\ n_\mu(-\omega, -q) \Pi_{\mu \nu} (\omega, q) n_\nu (\omega, q).
\end{multline}
The kernel $\Pi_{\mu \nu}$ appearing in the second term contains the quasiparticle contribution to the conductivity, $\sigma_{ij}^{qp}$,
\begin{equation*}
\sigma_{ij}^{qp} = \lim_{\omega \rightarrow 0} {1 \over i \omega} \langle j_i(- \omega, 0) j_j(\omega, 0)\rangle  = \lim_{\omega \rightarrow 0} {1 \over i \omega} \Pi_{ij}(\omega,  q=0),
\end{equation*}
where $j_i(\omega, q) = {\delta S_{\rm matter} \over \delta n_i(-\omega, -q)}$ is the quasiparticle current operator. Computing the DC conductivity from 
(\ref{eqn:eff-Action-LCS+A}), we find:
\begin{equation}
\label{eqn:LCS+matter-conductivities}
\sigma_{ij} = \frac{1}{2\pi}\lim_{\omega\rightarrow 0}
\epsilon_{ik} \epsilon_{jl} ({{k}\epsilon_{kl}+ 2\pi \sigma^{\rm qp}_{kl}})^{-1}.
\end{equation}
This implies that $\rho_{xy}=-\rho_{yx} = k$ while
$\rho_{xx} = 2\pi \sigma^{\rm qp}_{yy}$ and
$\rho_{yy} = 2\pi \sigma^{\rm qp}_{xx}$.
Thus, we see that one of the most remarkable features
of the experimental results in Ref. \onlinecite{Xia11}
has a natural explanation in our model:
$\rho_{xy}$ remains quantized while $\rho_{xx},\rho_{yy}$
can be temperature-dependent if $\sigma^{\rm qp}$ is diagonal.
Secondly, we note that the anisotropy in the DC resistivity comes
entirely from the induced anisotropy in the quasiparticle kinetic energy. 
By contrast, the transport due to the fluctuations in $S_{LCS}$
showed frequency-dependent anisotropy that resulted from subleading
terms in the gauge field action \cite{Mulligan10}.  So there is additional anisotropy in the
AC transport that is not present in the DC transport.  
In particular, AC transport shows low frequency conductivity that vanishes linearly and cubically along the two orthogonal directions.
The two types of
anisotropy come from different physical mechanisms -- anisotropy in
the gauge field kinetic energy versus anisotropy in the quasiparticle
kinetic energy -- although the ultimate cause is the same.

It remains to calculate $\Pi_{\mu \nu}$.  We summarize the calculation of $\Pi_{ii}$ for spatial $i$ below.  We introduce dissipation by assuming the quasiparticles have an elastic scattering lifetime equal to $\tau$ and single-particle gap $\Delta/2$.  Due to the anisotropy introduced by $\langle e_x \rangle$ in $S_{\rm matter}$, the longitudinal current-current correlation functions along the two spatial directions are related,
\begin{eqnarray*}
\left\langle {j_x}({\omega_n},0) {j_x}(-{\omega_n},0) \right\rangle & = &
(1+u \langle e_x \rangle^2)^{{1 \over 2}} f(i{\omega_n},T)\cr
\left\langle {j_y}({\omega_n},0) {j_y}(-{\omega_n},0) \right\rangle & = &
(1+u \langle e_x \rangle^2)^{-{1 \over 2}} f(i \omega_n, T),
\end{eqnarray*}
where
\begin{multline}
f(i{\omega_n},T) ={T \over \pi} \sum_{m} \int dq q^3
G (i\omega_{n+m},q) G(i\omega_{m},q)
\end{multline}
and after rescaling $q_x$ to  obtain the rotationally invariant form,
\begin{equation}
G^{-1} (i\omega_{m},q) = {i\omega_m - \Delta/2 -
q^2 + \frac{i}{\pi\tau} \text{Arg}(\Delta/2-i{\omega_n})}.
\end{equation}
Here, we use the fact that the imaginary part of the correlation
function (which gives the real part of the conductivity) is cutoff
independent so that the rescaling of the cutoffs can be neglected.  The diamagnetic contribution to the sum vanishes.

Replacing the sum over Matsubara frequencies, $\omega_m = 2 \pi T$, by a contour
integral, we have
${\rm Im} f(\omega + i \delta, T) = {\pi \over 4} \omega T \tau\, e^{-\Delta/2T},$
where we have made use of the large $\tau$ limit.  
Therefore, the longitudinal quasiparticle DC conductivities,
\begin{equation}
\label{eqn:conduct}
\sigma^{\rm qp}_{xx,yy} =
\mbox{$\frac{\pi}{4}$}(1+u{\left\langle {e_x} \right\rangle^2})^{\pm 1/2}\,
T\tau \,e^{-\Delta/2T},
\end{equation} 
where the $+$ $(-)$ refers to $\sigma_{xx}^{\rm qp}$ ($\sigma_{yy}^{\rm qp}$).
Inserting these expressions into (\ref{eqn:LCS+matter-conductivities}), we find that
\begin{equation}
\label{eqn:difference}
\rho_{xx} - \rho_{yy} \approx  \mbox{$\frac{\pi}{4}$}\,
u{\left\langle {e_x} \right\rangle^2} T\tau \,e^{-\Delta/2T} + {\cal O}(e^{- \Delta/T}).
\end{equation}
(\ref{eqn:conduct}) and (\ref{eqn:difference}) are assumed to be valid at temperatures $T < \Delta/2$, but high enough such that variable-range hopping can be ignored.
Thus, we have demonstrated theoretically the existence of a FQHE that has both anisotropic zero temperature AC transport as well as anisotropic finite temperature DC transport.

\label{fig:resistivityplotone}
%%%%%%%%%%%%%%%%%%%%%%%%%%%%%%%%
\begin{figure}[hb]
\begin{center}
\includegraphics[width=0.4\textwidth]{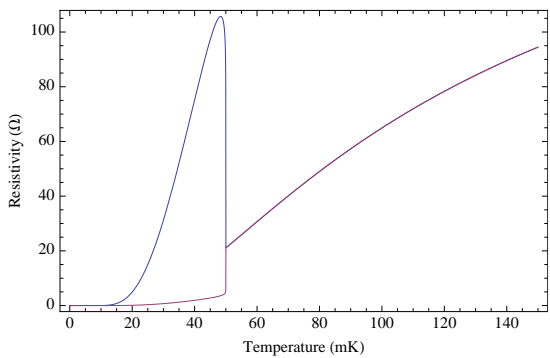}
\end{center}
\caption{Longitudinal resistivities $\rho_{xx,yy}$ along the easy x-axis (red) and hard y-axis (blue) obtained from the conductivities in (\ref{eqn:conduct}) are plotted versus temperature. 
The microscopic parameters entering the expressions in (\ref{eqn:conduct}) are found phenomenologically from the resistances measured in \cite{Xia11} as explained in the main text.}
\label{fig:resistivityplotone}
\end{figure}
%%%%%%%%%%%%%%%%%%%%%%%%%%%%

\label{fig:resistivityplottwo}
%%%%%%%%%%%%%%%%%%%%%%%%%%%%%%%%
\begin{figure}[hb]
\begin{center}
\includegraphics[width=0.4\textwidth]{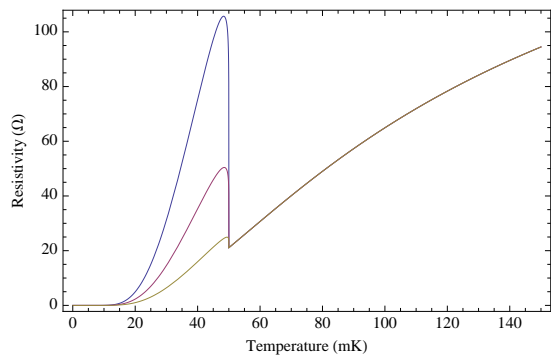}
\end{center}
\caption{Longitudinal resistivity $\rho_{yy}$ along the hard y-axis for three separate values of the parameter $m = u \langle e_x \rangle^2/(-t)^{1/6}$.  From top to bottom (blue, red, yellow), $m = 50, 10, 1$.
}
\label{fig:resistivityplottwo}
\end{figure}
%%%%%%%%%%%%%%%%%%%%%%%%%%%%

The precise temperature dependence of the DC resistivity is
determined by the behavior of $\langle e_x \rangle$ and the quasiparticle scattering time $\tau$.  
At temperatures near the rounded finite-temperature
phase transition, $\langle e_x \rangle$ can be identified with the order parameter of the
particular finite-temperature phase transition.
We expect this classical phase transition to be described by a theory lying on the Ashkin-Teller half-line or equivalently, the moduli space of the $c=1$ $\mathbb{Z}_2$ orbifold theory.
All theories along the line possess a global $\mathbb{Z}_4$ symmetry and an order parameter for the $\mathbb{Z}_4$-broken phase with critical exponent $1/16 < \beta < \infty$ \cite{criticalexponents} \cite{nematicLG}.
Since the Kosterlitz-Thouless critical point lies at the boundary point of this half-line, we expect the particular critical theory governing the transition to be largely determined by the degree of $SO(2)$ rotation symmetry-breaking in the experimental system.
Since the in-plane field appears to be a weak
symmetry-breaking field, as discussed in the Introduction, we
expect the transition to be fairly sharp.

For definiteness below, we take the transition to be in the universality class of the critical four-state Potts model which lies at the boson radius $r = 1/\sqrt{2}$ on the orbifold line.
In Fig.\ref{fig:resistivityplotone} are plotted the resistivities, $\rho_{xx, yy}$.
The order parameter for this transition $\langle {\cal O} \rangle \sim \langle e_x \rangle \sim (- t)^{1/12}$, where $t = (T - T_c)/T_c$ \cite{wu}.
(There is not a significant qualitative difference in the plots if the system has the full $SO(2)$ rotation symmetry so we have suppressed a separate discussion of the KT transition.)  

The microscopic parameters in (\ref{eqn:conduct}) are determined using the resistance measurements performed in \cite{Xia11} as follows.
For simplicity, we assume rotationally invariant relations between resistivities and resistances, $\rho_{xx} = f(L_x, L_y) R_{xx}$ and $\rho_{yy} = f(L_x, L_y) R_{yy}$, where $f(L_x, L_y)$ is some function depending on the sample lengths $L_{x,y}$ .
In other words, we ignore possible geometrical enhancements that may be present in translating between these two sets of quantities.
(See \cite{simon} for a discussion of the importance of this distinction in the context of anisotropic transport in $\nu = 9/2,11/2, ...$ half-filled Landau levels \cite{lilly1}.)
The  measured temperature dependences of $\rho_{xx}$ and $\rho_{yy}$ at zero in-plane field are fit well by Arrhenius plots, $\rho_{xx, yy} = A \exp(- \Delta / 2 T)$, with $\Delta = 225 {\rm mK}$ and $A = 10^{-2} h/e^2$.
We continue to use these values in the anisotropic regime, which we identify with the region $\bar{r} < 0$, when the in-plane field is of sufficient magnitude.
The temperature-independent value of $A$ over the temperature range $50 {\rm mK} < T < 150 {\rm mK}$ implies a quasiparticle scattering lifetime $\tau \sim {200 \over T}$.
An estimate of $5.6 \times 10^{-3} h/e^2$for the maximum value of $R_{xx}$ observed at a tilt angle of $66^\circ$ and achieved as $T \rightarrow 15 {\rm mK}$ from above implies $u  \langle e_x \rangle^2 \sim 50 (- t )^{1/6}$ with $T_c = 50 {\rm mK}$. 
We stress that the fitting of parameters used to obtain Fig.\ref{fig:resistivityplotone} is meant to be as optimistic as possible so as to determine the microscopic parameters of our theory if it is to apply to the experiment.

The height of the peak observed in Fig.\ref{fig:resistivityplotone} depends sensitively on the temperature-independent value of $(u \langle e_x \rangle^2)/(-t)^{1/6} = (u r)/((\lambda - \alpha)^2(-t)^{1/6})$.
In Fig.\ref{fig:resistivityplottwo}, we plot $\rho_{yy}$ for three different values of this parameter starting with the value used in Fig.\ref{fig:resistivityplotone}.
As the figure indicates, the peak decreases as this parameter is lowered.  

\label{fig:resistivityplotthree}
%%%%%%%%%%%%%%%%%%%%%%%%%%%%%%%%
\begin{figure}[hb]
\begin{center}
\includegraphics[width=0.4\textwidth]{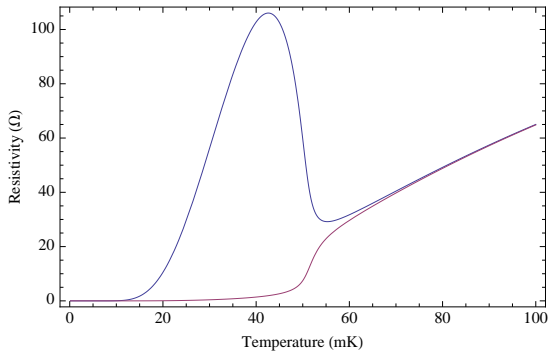}
\end{center}
\caption{In-plane field rounded longitudinal resistivities $\rho_{xx,yy}$ along the easy x-axis (red) and hard y-axis (blue) obtained from the conductivities in (\ref{eqn:conduct}) are plotted versus temperature.
A mean field crossover function has been used in the expression for the resistivities. 
}
\label{fig:resistivityplotthree}
\end{figure}
%%%%%%%%%%%%%%%%%%%%%%%%%%%%

We expect the in-plane magnetic field $B_{||}$ to act as a small symmetry-breaking field on this finite temperature transition. This will lead to a rounding of the resistivity curves in Fig.\ref{fig:resistivityplotone}. The order parameter now behaves as
\begin{equation}
\label{eqn:rounding}
\langle e_x \rangle \sim B_{||}^{1/ \delta} g_{\pm}\Big({(\pm t)^{\beta} \over B_{||}^{1/\delta}}\Big),
\end{equation} 
where the $\pm$ is determined by the sign of $t$, and the critical exponents $\beta = 1/12$ at the four-state Potts point and $\delta = 15$ along the orbifold line.
Integral expressions for the scaling functions $g_{\pm}$ are known \cite{Lukyanov:1996jj}; 
a precise functional form, however, is not. 
Scaling dictates that $g_{-}(x = 0) = g_{+}(x=0)$ are finite and non-zero,
$g_{-}(x) \sim x$ as $x \rightarrow \infty$, and $g_{+}(x) = 0$ for $x > x_{\rm crit} \sim 1$.

Since we do not have an explicit functional form for $g_{\pm}$, let us simply model the transition using mean field theory in order to obtain a picture for the rounding of the transition. 
(We do not mean to imply that the crossover function for the $\mathbb{Z}_4$ transition is in any way similar $\phi^4$ mean field crossover function. Rather, we only want a picture for how the transition might be rounded.)
The $\phi^4$ mean field critical exponents $\beta = 1/2$ and $\delta = 3$. 
Specification of the free energy, $F = {1 \over 2} T_c t \phi^2 + {1 \over 4} \phi^4 - \phi h$, allows the calculation of $g_{\pm}$ via minimization of $F$ with respect to $\phi$.
We select the root,
\begin{equation}
\label{eqn:mean}
\langle \phi \rangle  =  h^{1/3} \Big({- 2(3)^{1/3} x^2 + 2^{1/3} (9  + \sqrt{81 + 12 {\rm sgn}(t) x^6})^{2/3} \over 6^{2/3}(9  + \sqrt{81 + 12 {\rm sgn}(t) x^6})^{1/3})}  \Big),
\end{equation}
where $x = |T-T_c|^{1/2}/h^{1/3}$. 
It  satisfies the scaling requirements detailed in the previous paragraph.
Substituting $u \langle e_x \rangle^2 = 7 \langle \phi \rangle^2$ at $h=1$ into our expressions for the resistivities using the same values for the overall scale of the resistivity and behavior of the scattering time $\tau$ as above, we find Fig. 3.

When $r \geq 0$, the form (\ref{eqn:LCS+matter-conductivities}) and (\ref{eqn:conduct})
of the finite temperature DC conductivity matrix still holds.  However, $\langle e_x \rangle$ is zero and the longitudinal conductivity along the two directions coincides.  Note that non-zero AC conductivity at the $r=0$ critical point requires disorder exactly like the $r<0$ regime \cite{Mulligan10}.

\paragraph{Discussion.}

In this paper, we have given an explanation of one of the most striking
aspects of the data of Ref. \onlinecite{Xia11}: the anisotropy of the longitudinal
resistances coexisting with quantized Hall resistance. Our theory further predicts
that, while one of the resistances will increase with decreasing temperature at temperatures just below the (rounded) finite-temperature phase transition
at which nematic order develops, as observed \cite{Xia11},
both longitudinal resistances will, eventually, go to zero at the lowest
temperatures, which is yet to be observed. Transport beyond the linear
regime, the nature of the massive quasiparticles in the anisotropic phase, and a more complete determination of the values of the parameters in the effective Lagrangian in terms of microscopic variables are interesting open problems.

\acknowledgments
We thank J. Chalker, J.~P. Eisenstein, E. Fradkin, S. Kivelson, H. Liu, J. McGreevy, S. Shenker,  S. Simon and J. Xia for helpful discussions, and thank the Aspen Center for Physics for hospitality.  M. M. acknowledges the hospitality of the Stanford ITP, the Galileo Galilei ITP and INFN, and Oxford University while this work was in progress.  M. M. was supported in part by funds provided by the U. S. Department of Energy (D. O. E.) under cooperative research agreement DE-FG0205ER41360.
C.N. was supported in part by the DARPA-QuEST program.

\bibliography{qhbiblio}

\begin{thebibliography}{21}
\bibitem[{CS-()}]{CS-normalization}
\bibinfo{note}{One can rewrite this action in terms of ${\tilde a}_\mu =
  \nu^{-1} a_\mu$ so that the Chern-Simons term will have an integer
  coefficient $\nu^{-1}$ and the theory can be defined on arbitrary manifolds.}
\bibitem[{cri()}]{criticalexponents}
\bibinfo{note}{The critical exponent $\beta$ of the order parameter field along
  the $c=1$ orbifold line can be determined using the relation $\beta = \nu
  (\eta / 2)$ between critical exponents. $\nu = 2 - x$ where $x$ is the
  scaling dimension of the `energy' operator of the CFT. $1/2< \nu < \infty$ as
  the boson radius $\infty > r > 1/\sqrt{2}$. The twist operator of the CFT has
  scaling dimension fixed at $\eta/2 = 1/8$ along the critical line (since it
  simply acts to change boundary conditions). The critical four-state Potts
  model and $Z_4$ parafermion CFT have $\beta = 1/12$ and $3/32$, respectively,
  while $\beta$ formally diverges at the KT point where the correlation length
  diverges as $\exp((T/T_c - 1)^{-1/2})$ as the critical temperature is
  approached. The exponent characterizing the decay of the order parameter at
  the critical point with respect to a symmetry-breaking field, $\delta =
  (4-\eta)/\eta = 15$, along the critical line.}
\bibitem[{nem()}]{nematicLG}
\bibinfo{note}{Note that a conventional Landau-Ginzburg description of the
  classical phase transition between a nematically ordered and isotropic state
  is described by a free energy functional of a symmetric, traceless $n\times
  n$ matrix $N_{\alpha \beta}$. Landau-Ginzburg theory predicts a continuous
  transition when $n=2$ while a first-order transition is expected when $n \geq
  3$ because of an allowed cubic term in the free energy. The
  $\mathbb{Z}_4$-invariant critical point we study describes a continuous
  symmetry-breaking transition with vector order parameter $\langle e_i
  \rangle$. It is the $\mathbb{Z}_2$-invariant director $\langle e_i \rangle^2$
  that enters physical quantities like the resistivity and allows the
  description of a continuous nematic transition in terms of certain response
  functions.}
\end{thebibliography}

\paragraph{Appendix.}

In this Appendix, we show
that, as a result of the violation of the conditions of Kohn's theorem, the location of the
cyclotron pole can vary as $B_{||}$ is increased.  
(It is also possible that additional spectral weight shows up at ${\cal O}(q^2)$,
but we shall not study this possibility in any detail.)
We do this by identifying the leading pole in the density-density
response with the gap between the lowest and first-excited states in the center-of-mass part of the quantum mechanical many-body wave function.  
This identification is correct for vanishing in-plane field $B_{||}$ and we believe it holds for perturbatively small values of $B_{||}$ as well, where separation of variables into
center-of-mass and relative coordinates is well-defined.
Of course, this simple example can, at best, give us a few clues
about the real system, which is far more complicated. One of these is that 
the pole moves towards the origin (at least initially) as $B_{||}$ is increased from zero.  
This justifies our study of the model (\ref{CSLG}) with varying $\bar{r}$.

We begin with the quantum mechanics problem of two mutually interacting three-dimensional electrons in a background magnetic field.  
This can be easily generalized to an arbitrary number of particles.
We take their motion along the $x-y$ plane to be unconstrained but
subject to a confining potential along the $z$-direction.  
The Hamiltonian is  
\begin{eqnarray}
\label{hami}
H & = & \sum_{i = 1,2} \Big[{1 \over 2 m_e} \Big(\partial_{x_i}^2 + (i \partial_{y_i} -  (B x_i + B_{||} z_i))^2 - \partial_{z_i}^2  \Big) \cr 
& + & A_0(z_i) \Big] + V(|{\bf x}_1 - {\bf x}_2|),
\end{eqnarray}  
where ${\bf x}_i = (x_i, y_i, z_i)$ labels the position of the two particles.  
(The gauge chosen for the vector potential is consistent with a spatial geometry that is of a finite length along the $x,z$-directions, and infinite along the $y$-direction.  
Note, however, we are essentially ignoring the finite length along the $x$-direction in the discussion below so we can think of it as being large compared to the length scale provided by the confining potential along the $z$-direction.)

We consider the component of the magnetic field lying along the $x$-direction
to be a perturbation to the system.  
It is convenient to switch coordinates to the center-of-mass and relative coordinate frame.  
Choosing
\begin{equation}
\label{centerrel}
{\bf X} = {1 \over 2}({\bf x}_1 + {\bf x}_2), \quad \rho = {\bf x}_1 - {\bf x}_2,
\end{equation}
the Hamiltonian becomes
\begin{eqnarray}
\label{hamcenrel}
H & = & {1 \over 2 (2 m_e)} \Big(-\partial_{X_x}^2 + (i \partial_{X_y} - 2 (B X_x + B_{||} X_z))^2 - \partial_{X_z}^2  \Big)  \cr
 & + & {1 \over 2 (m_e/2)} \Big(-\partial_{\rho_x}^2 + (i \partial_{\rho_y} - {1 \over 2} (B \rho_x + B_{||} \rho_z))^2 - \partial_{\rho_z}^2  \Big)  \cr
 & + & \sum_{\pm} A_0(X_z \pm {1 \over 2} \rho_z)   + V(|{\bf \rho}|).
\end{eqnarray}
Aside from the confining potentials $A_0(z_{1,2}) = A_0(X_z \pm {1 \over 2} \rho_z)$, the center-of-mass and relative coordinates are decoupled.  
%(When there is no in-plane magnetic field, it is best to take a mixed coordinate frame with center of mass and relative coordinates along the $x-y$ plane and to leave the $z_i$ coordinates untouched.)

At $B_{||} = 0$,  motion in the $z$-direction decouples from the motion in the plane and we are left with a collection of two-dimensional electrons indexed by their band or energy along the $z$-direction.  (Here, we are assuming the pair potential only depends on the separation of the electrons in the $x-y$ plane; the well width is assumed small compared to the magnetic length $B^{-1/2}$. This is not
the case in the experiments of Refs. \onlinecite{Xia10,Xia11}, so the violations of Kohn's
theorem will be larger than in our simple model.)
Given a $z$-eigenfunction, the center-of-mass part of the wave function executes oscillatory motion at the cyclotron frequency $2 B/ (2m_e)$.
This is the generalization of Kohn's theorem to the situation where electrons are confined along the direction parallel to the magnetic field.
When the spacing between the energy levels of the $z_i$-eigenfunctions greatly exceeds the cyclotron frequency, it is possible to ignore higher sub-bands when considering low-energy properties of the system.  However, this is not the case in the experiments of Refs. \onlinecite{Xia10,Xia11}.  
%The spectrum of the relative part of the Hamiltonian is influenced by the inter-particle potential $V$.  For example, if we consider Coulomb interactions, $V|\rho| = e^2/(\epsilon |\rho|)$, with dielectric constant $\epsilon$, we can estimate the interaction energy as $E_{{\rm int}} \sim e^2 \sqrt{e B}/\epsilon$.

Now consider $B_{||} \neq 0$.
There is now a direct mixing between motion in the $z$-direction and motion in the plane.  
This mixing mediates a coupling at higher orders in $B_{||}$ between the planar center-of-mass and relative degrees of freedom.
%(It is clear from (\ref{hamcenrel}) that for general $A_0$, there is no good choice of coordinate system such that the wave function factorizes into center-of-mass and relative factors.)  
Thus, there there is no requirement of a pole at the cyclotron frequency in the density-density correlator.  
This follows from the fact that the full three-dimensional Galilean symmetry (except for $X_y, \rho_y$ translations) is broken when there is {\it both} an in-plane field and a non-zero confining potential along the direction normal to the plane.  
If either the confining well or in-plane field are removed, there will be a Kohn pole at $\omega_c$.

We would like to better understand departures of the pole from the cyclotron frequency in this more general situation with non-zero in-plane field.  
Namely, we would like to know how the location of the pole varies with $B_{||}$.  
We can obtain some intuition by studying a special case for the form of the confining potential.  Take the confining potential to be quadratic, $A_0(z) = {\lambda^2 \over 2} z^2$.  Then, because 
\begin{equation}
\label{quadpot}
A_0(X_z + {1 \over 2} \rho_z) + A_0(X_z - {1 \over 2} \rho_z) = \lambda^2 (X_z^2 + {1 \over 4} \rho_z^2),
\end{equation}
the center-of-mass and relative coordinate motion are still decoupled.
This decoupling is not generic; a quartic potential, for example, couples $X_z$ and $\rho_z$ together.
However, we will argue that some conclusions drawn from the quadratic case are general.

We know that at $B_{||} = 0$, the Kohn pole corresponds to the splitting between the ground and first excited states of the center-of-mass motion.  The relative coordinate is irrelevant both when $B_{||} = 0$ and for a quadratic electric potential, and so we drop it from our discussion.  Thus, the Hamiltonian we study perturbatively in $B_{||}$ is
\begin{equation}
\label{pertham}
H = H_0 + H_1,
\end{equation}
where
\begin{eqnarray}
\label{perthamdef}
H_0 & = & {1 \over 2 (2 m_e)} \Big(-\partial_{X_x}^2 + (i \partial_{X_y} - 2 B X_x)^2 - \partial_{X_z}^2  \Big) \cr 
& + & \lambda^2 X_z^2, \cr
H_1 & = &  {B_{||} B \over m_e} (i \partial_{X_y} - 2 B X_x) X_z + {\cal O}(B_{||}^2).
\end{eqnarray}
First, we note that translation invariance along the $X_y$-direction allows us to replace derivatives with respect to $X_y$ with the momentum $k_y$ along this direction.  Next, we shift the $X_x$ coordinate by defining $\tilde{X}_x = {k_y c \over 2 e B} - X_x$.  The Hamiltonian has the form,
\begin{eqnarray}
\label{shiftham}
H & = & {1 \over 4m_e} \Big( - \partial_{\tilde{X}_x}^2 + 4B^2 \tilde{X}_x^2 - \partial_{X_z}^2 + 4 m_e \lambda^2  X_z^2 \Big) \cr
& + & {B_{||}  B \over m_e} \tilde{X}_x X_z + {\cal O}(B_{||}^2),
\end{eqnarray}
where terms proportional to $B_{||}$ are taken to be a perturbation.  

Our goal is to determine the spectral flow as a function of $B_{||}$ of the ground and first excited energy levels.
Dividing out by the irrelevant $X_y$ factor (which determines the degeneracy of the Landau levels in a rectangular sample, but is inconsequential here), the eigenfunctions of the above coupled harmonic oscillator Hamiltonian at $B_{||} = 0$ take the form:
\begin{eqnarray}
\label{eigfuc}
\Psi(\tilde{X}_x, X_z)_{m,n} & = & c_{m,n}\exp(- {M \omega_c \over 2} \tilde{X}_x^2) \exp(- {M \omega_z \over 2} X_z^2) \cr
& \times & H_m( \sqrt{M \omega_c} \tilde{X}_x) H_n(\sqrt{M \omega_z} X_z),
\end{eqnarray}
where $\omega_c = B/m_e$, $\omega_z = \lambda/\sqrt{m_e}$, $M = 2 m_e$,  $H_n(X)$ denote Hermite polynomials, and the $c_{m,n}$ are normalization constants.  We assume that $\omega_c < \omega_z < \infty$.  

(For electrons moving in a Ga-As quantum well at $\nu = 7/3$, we can estimate $\omega_c$ and $\omega_z$.  
Given a band mass $m_e \sim .07 m_f$, where $m_f$ is the free electron mass and transverse magnetic magnetic field of $2.82$ T, we estimate an $\omega_c \sim 3 \times 10^{-3} {\rm eV} \sim 30 {\rm K}$.
$\lambda$ has engineering dimension equal to $[{\rm Mass}]^{3/2}$ so we take it to be proportional to $1/w^{3/2}$, where $w$ is the well width which is 40 nm for the experiment in \cite{Xia11}.
We fix the order of the proportionality constant via the estimate of the Landau level sub-band gap given in Fig. 2 of \cite{Xia10}.  
We find $\omega_z \sim x/w^{3/2}$, where $x = 10^{-3}$.
Thus, $\omega_z/\omega_c \sim 2$.
As the filling fraction is lowered, the ratio $\omega_z/\omega_c$ is increased and so the discussion below becomes less relevant as the two scales are too far apart.
Note also that this ratio approaches unity as the proportionality constant between the band and free electron masses is lowered.)

The perturbative shift in the energy of a state to second order in $H_1$ is given by the formula,
\begin{equation}
\label{energy}
E_{m,n} = E_{n,m}^{(0)} + \langle m, n| H_1 |m,n\rangle + \sum_{|k\rangle \neq |m,n\rangle} {|\langle k| H_1 |m,n\rangle|^2 \over E^{(0)}_{m,n} - E^{(0)}_k},
\end{equation}
where $E^{(0)}_{m,n}$ is the unperturbed energy of the state $\Psi(\tilde{X}_x, X_z)_{m,n} := |m,n\rangle$.
We are interested in the difference $E_{0,0} - E_{1,0}$.  At zeroth-order, this difference is equal to the cyclotron frequency, $\omega_c$.  The first-order term on the RHS of (\ref{energy}) vanishes because the perturbation is linear in both $\tilde{X}_x$ and $X_z$ (the ground state is nodeless).  Now consider the second-order term.  The state (or collection of states since we are ignoring the $k_y$ dependence) mixed with the ground state $|0,0 \rangle$ by the perturbation is $|1,1\rangle$, while $|0, 1\rangle$ is the state mixed with $|1, 0 \rangle$.  Because the wave functions factorize, 
\begin{equation}
\label{equals}
|\langle 1, 1| H_1 | 0, 0 \rangle| = |\langle |0, 1| H_1 | 1, 0 \rangle|,
\end{equation}
 however, 
 \begin{equation}
 \label{energyden}
|E^{(0)}_{0,0} - E^{(0)}_{1,1}| = \omega_c + \omega_z  > \omega_z - \omega_c = |E^{(0)}_{1,0} - E^{(0)}_{0,1}|,
 \end{equation}
 with both energy denominator differences being negative.  So while both $E_{0,0}$ and $E_{1,0}$ are shifted downwards, the ground state is shifted less than the first excited state because of the difference in magnitude of the energy denominators.  This implies that the location of the would-be Kohn pole is decreased from the cyclotron frequency.  
(There is no contradiction with general level repulsion expectations as we are studying a systems with more than two states.) 
Notice that this result requires mixing between the different $X_z$-bands and is not present if we take the gap between these energy levels to infinity.  Contributions from other excited states only occur at higher orders in perturbation theory.  Note also that we dropped a term in the perturbing Hamiltonian quadratic in $B_{||}$ and so it could, in principle, compete at the same order as the second-order result above.  However, this term has no consequence on the energy difference as it shifts both energies by the same amount.

The above analysis implies that the location of the leading pole in a small momentum expansion of the density-density correlator moves towards the origin as an in-plane field is applied.  
This conclusion was drawn using a certain form of confining potential in the direction transverse to the $x-y$ plane. How general are these results? If we consider a more
general form of the confining potential,
there will be a coupling between the center-of-mass and relative degrees of freedom.
Nevertheless, for small $B_{||}$, our results hold generally. In this limit,
we can ignore the coupling between the center-of-mass and relative degrees of freedom.
The perturbation couples the planar and $z$ motion of the center-of-mass at leading order
while the coupling between the planar center-of-mass and relative degrees of freedom only occurs at higher order in perturbation theory in $B_{||}$.
Although the $X_z$ eigenfunctions will take a different functional form and the spacing between these eigenfunctions will no longer be in regular multiples of $\omega_z$, the above argument goes through unchanged as long as the gap between the lowest and first-excited $X_z$ eigenfunction is greater than $\omega_c$. If $B_{||}$ is not small, then we cannot ignore
the coupling between the center-of-mass and relative degrees of freedom.
This can further modify the distribution of spectral weight,
but we do not have any simple argument for whether this
coupling will move the pole, broaden the pole
into a Lorentzian, or change its spectral weight.
At any rate, we can say that the coupling between the center-of-mass
and relative degrees of freedom will almost certainly cause further deviations
from expectations based on Kohn's theorem.
Happily, the effective field theory (\ref{CSLG}) describes a system where the would-be Kohn pole is different from the bare cyclotron frequency through the variation of $\bar{r}$;
we pursue its study in the body of the paper.

%\insert{qhbiblio.bbl}

\end{document}